\documentclass[twocolumn,preprintnumbers,amsmath,amssymb,prb,floatfix,superscriptaddress]{revtex4-2}

\usepackage{graphicx,tabularx}
\usepackage[version=4]{mhchem} 
\usepackage[T1]{fontenc}       
\usepackage{graphics}

\begin{document}
\title{Linker Functionalization in MIL-47(V)-R Metal-Organic Frameworks: Understanding the Electronic Structure}
\author{Danny E. P. Vanpoucke}
\email{Danny.Vanpoucke@UHasselt.be}
\affiliation{UHasselt, Institute for Materials Research (IMO-IMOMEC), Agoralaan, 3590 Diepenbeek, Belgium}
\affiliation{IMOMEC, IMEC vzw, 3590 Diepenbeek, Belgium}


\noindent This work was published as:\\
\textit{J. Phys. Chem. C}  \textbf{121(14)}, 8014-8022 (2017)\\
doi: \url{10.1021/acs.jpcc.7b01491}\\

\begin{figure}[!h]
  \centering
  \includegraphics[width=8cm,keepaspectratio=true]{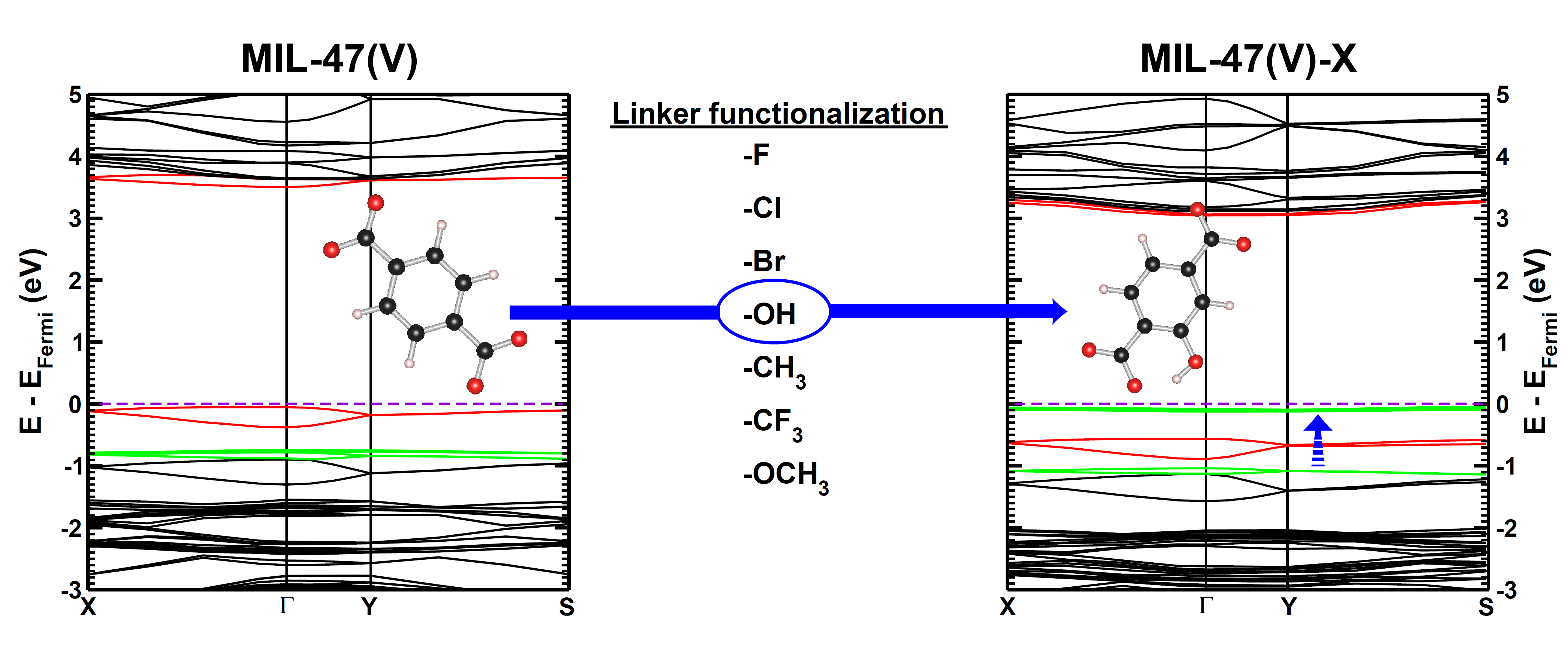}
\end{figure}

\begin{abstract}
Metal-Organic Frameworks (MOFs) have gained much interest due to their intrinsic tunable nature. In this work, we study how linker functionalization modifies the electronic structure of the host MOF, more specifically the MIL-47(V)-R (R=\ce{-F}, \ce{-Cl}, \ce{-Br}, \ce{-OH}, \ce{-CH3}, \ce{-CF3}, and \ce{-OCH3}). It is shown that the presence of a functional group leads to a splitting of the $\pi$-orbital on the linker. Moreover, the upward shift of the split-off $\pi$-band correlates well with the electron withdrawing/donating nature of the functional groups. For halide functional groups the presence of lone-pair back donation is corroborated by calculated Hirshfeld-I charges. In case of the ferromagnetic configuration of the host MIL-47(V$^{\mathrm{+IV}}$) material a half-metal to insulator transition is noted for the \ce{-Br}, \ce{-OCH3}, and \ce{-OH} functional groups, while for the anti-ferromagnetic configuration only the hydroxy-group results in an effective reduction of the band gap.
\end{abstract}
\maketitle

\section{Introduction}\label{Introduction}
\indent Metal-organic frameworks (MOFs) have, over the last few decades, emerged as a class of highly tuneable crystalline materials that show ultrahigh porosity (typically $50$\% and more) and very large internal surface areas (typically $1000$ to $10 000$m$^2$/g). These materials consist of two types of building blocks: (1) metal or metal-oxide centres connected through (2) organic linker molecules. The variability of these inorganic and organic components allows an almost infinite set of possible combinations, each with their own specific properties. As such, more than $20 000$ different MOFs have been reported in the Cambridge Structural Database.\cite{FurukawaH:Science2013} By specifically selecting the building blocks, MOFs can be designed with pores of predefined size and shape. This makes MOFs very attractive materials for applications: \textit{e.g.} gas storage,\cite{MurrayL:ChemSocRev2009, SzilagyiPA:FrontEnerRes2016, SpanopoulosI:JACS2016} carbon dioxide capture,\cite{BourrellyS:JACS2005, LiJianRong:ChemRev2012, LlewellynPL:JPhysChemC2013, FernandezM:JPhysChemLett2014} sensor applications,\cite{KhoshamanAH:SensActB2012} luminescence,\cite{AllendorfMD:ChemSocRev2009, HendrickxVanpoucke:InorgChem2015, BeukenVanpoucke:AngewChemIntEd2015} catalysis,\cite{YoonM:ChemRev2012} and many more. \cite{RosseinskyMJ:MicroMeso2004, LlewellynPL:AngewChemIntEd2006, AlaertsL:AngewChemIntEd2007, GasconJ:AngewChemIntEd2010, WangB:JACS2016}\\
\indent Also from the fundamental point of view, MOFs show many interesting properties. Certain MOFs exhibit specific magnetic and multiferroic properties.\cite{KurmooM:ChemSocRev2009, RogezG:AngewChemIntEd2010, CentroneA:Small2010, CanepaP:PhysRevB2013, WangZ:PhysRevB2013, ChenX:DaltonTrans2013, CortijoM:CrystGrowthDes2014, SibilleR:PhysRevB2014} Other porous MOFs show, under external stimuli (temperature, pressure or gas sorption), reversible flexible behavior, called breathing. This breathing leads to large variations of the pore size while maintaining the topology of the MOF.\cite{LlewellynPL:AngewChemIntEd2006, SerreC:AdvMater2007, LeclercH:JPCC2011, MurdockCR:CoordChemRev2014, ImJ:ChemMater2016} In addition, it was also recently found that the metal-oxide clusters of MOFs can undergo structural phase-transition, while retaining the framework topology, adding a new dimension to the possibilities of MOF-tuning.\cite{Platero-PratsA:JACS2016}\\
\indent In addition to varying the building blocks, MOF properties such as their sorption, selectivity and their stability can also be tuned in a systematic fashion through the functionalization of the organic linker molecules.\cite{ThorshaugKEspenFL:2013JPhysChemC, PhamHQ:JPhysChemC2014, HendrickxVanpoucke:InorgChem2015, SzilagyiPA:FrontEnerRes2016} In case of flexible MOFs such functionalization can influence their breathing characteristics.\cite{YotPG:EurJInorgChem2016}\\
\indent Although MOFs are typically insulators or wide band gap semiconductors, some research is also being directed to the creation of conducting MOFs. This is either done through filling of 1D channels with conducting polymers,\cite{DharaB:JPhysChemLett2016} by selective metal doping of the inorganic nodes,\cite{FuentesCabreraM:JChemPhys2005} or through discovery of intrinsic conducting or semiconducting MOFs.\cite{ParkSS:JAmChemSoc2015, SheberlaD:JAmChemSoc2014} For the more prevalent insulating MOFs, band gap tuning attracts much attention as it provides opportunities in the fields of luminescence and sensing.\cite{ThorshaugKEspenFL:2013JPhysChemC, HendonCH:JACS2013, HendrickxVanpoucke:InorgChem2015, LingS:JPhysChemC2015, YuD:DaltonTrans2012} The very stable UiO-topology is well studied,\cite{ThorshaugKEspenFL:2013JPhysChemC, ThorshaugKEspenFL:InorgChem2014, HendrickxVanpoucke:InorgChem2015} as well as the IRMOFs.\cite{PhamHQ:JPhysChemC2014}
Also the electronic structure of the MIL-47(V) MOF has received some attention. Meilikhov and co-workers studied how cobaltocene can be used to reduce V$^{IV}$ to V$^{III}$ in MIL-47(V).\cite{MeilikhovMAngewChemIntEd2010} This mixed valency version of MIL-47(V) is found to have an anti-ferromagnetic ground state, in which the separate chains have a ferromagnetic spin configuration. This differs significantly from the anti-ferromagnetic ground state found, by the present author, for MIL-47(V$^{IV}$), in agreement with single crystal magnetic susceptibility measurements.\cite{VanpouckeDannyEP:2014e_Beilstein, BartheletK:AngewChemIntEd2002} In the same work, it was also shown that the in-chain spin configuration could be linked to the breathing properties of the MIL-47(V), and that different spin configurations give rise to very different valence band edges. The flexibility of mixed oxidation state MIL-47(V$^{III}$/V$^{IV}$) was also investigated by Leclerc and co-workers, showing the relative rigidity of the MIL-47(V) originates from the V$^{IV}$ species.\cite{LeclercH:JPCC2011} In an experimental study, Centrone and co-workers showed that linker functionalization can modify the interactions between the V atoms by modifying the local crystal geometry around the V atoms.\cite{CentroneA:Small2010} In a recent computational study, Ling and Slater investigated the role of the metal center in the metal-oxide chains of MIL-53(X)/MIL-47(X) MOFs.\cite{LingS:JPhysChemC2015} They note there is a clear difference in the character of the band gap edges for magnetic and non-magnetic metal centers, with the conduction band edge showing rather a molecular character for the non-magnetic centers, while a metal character is found for the magnetic centers.\\
\indent In this work, the influence of electron drawing and electron donating functional groups on the electronic structure of the parent MIL-47(V$^{IV}$) framework is investigated. Modifications of the atomic structures are also briefly touched as is charge transfer due to functionalization. The focus goes to a series of isotypic MIL-47(V$^{IV}$)-R (R= \ce{-F}, \ce{-Cl}, \ce{-Br}, \ce{-CH3}, \ce{-CF3}, \ce{-OCH3}, and \ce{-OH}) in addition to unfunctionalized MIL-47(V$^{IV}$), and predictions are made on the behavior of other functional groups.\\

\section{Computational Details}
\begin{figure*}
  \includegraphics[width=15cm,keepaspectratio=true]{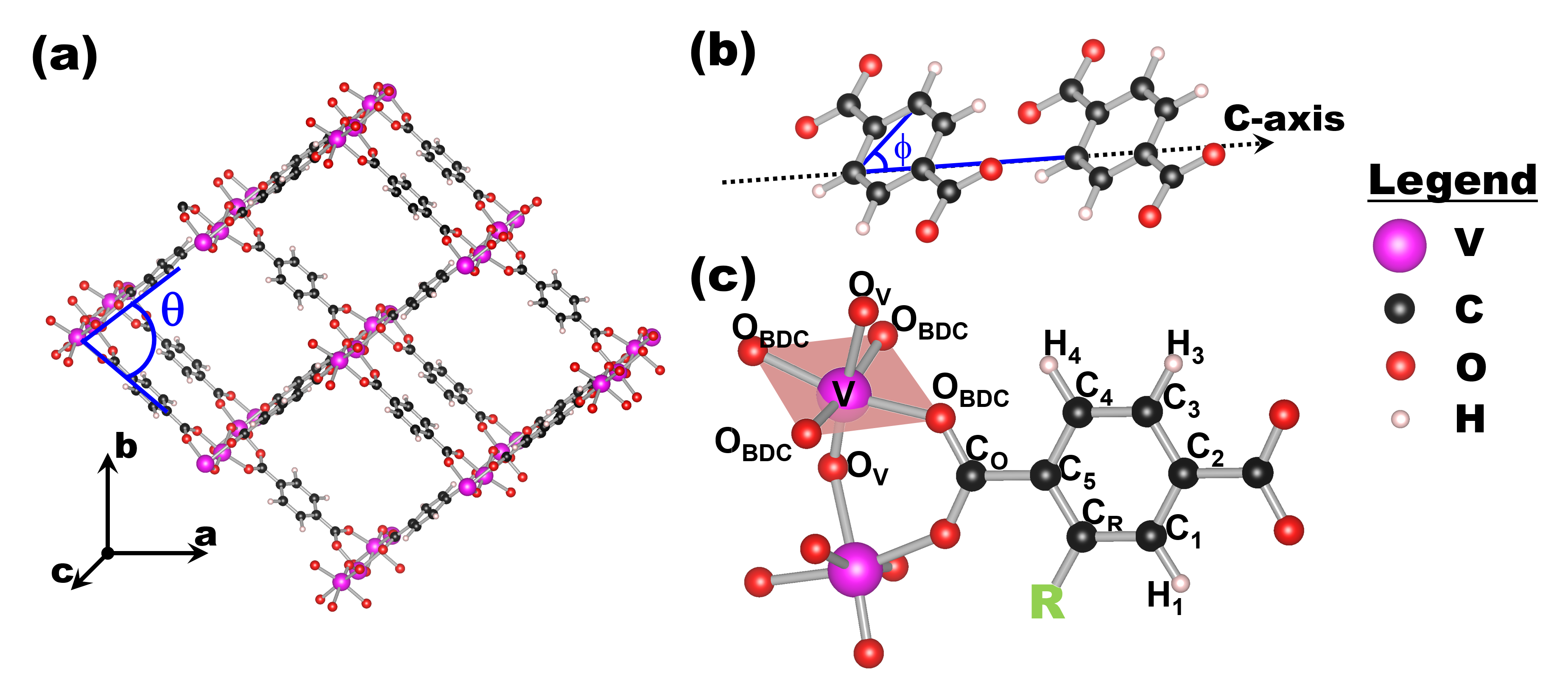}\\
  \caption{(color online) (a) Ball-and-stick representation of a super cell of the MIL-47(V) MOF. Pink, red, black, and white spheres indicate V, O, C and H
  positions, respectively. The MOF opening angle $\theta$ is indicated. (b) Representation of two periodic copies of a linker along the c-axis, indicating the
  linker rotation angle $\phi$. (c) Ball-and-stick representation of a single functionalized linker bound to the vanadyl chain. The position of the functional
  group is indicated as R (with R=\ce{-F}, \ce{-Cl}, \ce{-Br}, \ce{-OH}, \ce{-CH3}, \ce{-CF3}, and \ce{-OCH3}), as are the different atomic positions. For the
  system with two functional groups per linker (\ce{-2,5F}), the second functional group is placed at position H$_3$, making C$_3$ equivalent to C$_R$, C$_4$ to
  C$_1$, and C$_5$ to C$_2$. Ball-and-stick images were generated using VESTA.\cite{VESTA:JApplCryst2008}}\label{fig:MIL47ref}
\end{figure*}

\indent The atomistic model used in the density functional theory (DFT) calculations of the functionalized MIL-47(V$^{IV}$)-R (R= \ce{-F}, \ce{-Cl}, \ce{-Br}, \ce{-CH3}, \ce{-CF3}, \ce{-OCH3}, and \ce{-OH}) consists of a conventional unit cell containing four formula units (\textit{i.e.} containing $72$ to $88$ atoms). The general topology of the MOFs is given by the ball-and-stick representation in Fig.~\ref{fig:MIL47ref}a while the different atomic sites are shown in Fig.~\ref{fig:MIL47ref}c. Starting from the experimental lattice parameters and volumes,\cite{BiswasVanpoucke:JPhysChemC2013} the structures are optimized under the constraint of constant volume to allow for direct comparison to the experimental results. The conventional unit cell contains four V atoms, two on each vanadyl chain. The unpaired electrons of these four V atoms, one electron each, are considered both in the anti-ferromagnetic (AF) ground state configuration\cite{BartheletK:AngewChemIntEd2002} and the ferromagnetic (FM) configuration. The latter appears to be predominantly present in MIL-47(V) powder, as has been inferred from X-Ray Diffraction data\cite{BogaertsVanpoucke:CrystEngComm2015} and mercury intrusion experiments.\cite{YotPG:ChemSci2012, VanpouckeDannyEP:2014e_Beilstein} The different spin configurations are manually imposed during initialization of the structure optimization, and for each of the following steps it is verified that the imposed spin configuration is retained.\\
\indent The DFT calculations are performed with the VASP program,\cite{Kresse:prb93, Kresse:prb96, Kresse:prb99} using both the PBE, a generalized gradient approximation (GGA) functional, and hybrid HSE06 functionals.\cite{PBE_1996prl, HeydJ:JChemPhys2003_HSE,Heyd:JChemPhys2005_HSE_BG, GrimmeS:JPhysChemC2014_HSE06D3} Due to the extreme computational cost of hybrid functionals such as HSE06 ($\times 70$ compared to PBE for these systems), the latter is only used to obtain a very accurate picture of the electronic structure. Structure optimization is performed with the PBE functional, which is well-known to give good quality lattice parameters and mechanical properties for periodic materials. For comparative reasons, the electronic structure calculations were also performed at the PBE level. Atomic charges are calculated using the iterative Hirshfeld-I approach as implemented in our HIVE code.\cite{BultinckP:2007JCP_HirRef, VanpouckeDannyEP:2013aJComputChem, VanpouckeDannyEP:2013bJComputChem, HIVE_REFERENCE} A detailed description of the used computational parameters is given in Sec.~S1 of the SI. The optimized atomic structures can be obtained free of charge from The Cambridge Crystallographic Data Centre: CCDC $1507811$--$1507830$.\\

\begin{table*}[!tb]
  \caption{Structural parameters of functionalized MIL-47(V)-R MOFs with an AF spin configuration.$^{a}$}\label{Tab:MOFStruct_AF}
  \begin{tabular}{l|rrrrrr|rrrl}
    & \multicolumn{6}{|c|}{this work} & \multicolumn{4}{|c}{experiment}\\
    & $a$ & $b$ & $c$ & Vol & $\theta$ & $\phi$ & $a$ & $b$ & $c$ & \\
    & (\AA) & (\AA) & (\AA) & (\AA$^3$) & ($^{\circ}$) & ($^{\circ}$) & (\AA) & (\AA) & (\AA)&    \\
  \hline
MIL-47(V)  & $16.146$ & $14.040$ & $6.839$ & $1550.344$ & $82.0$ & $5.5$  & $17.434$ & $13.433$ & $6.620$ & [\onlinecite{BiswasVanpoucke:JPhysChemC2013}]\\
\hline
-F         & $16.269$ & $13.957$ & $6.862$ & $1558.092$ & $81.3$ & $6.3$  & $16.406$ & $13.824$ & $6.870$ & [\onlinecite{BiswasS:PCCP2013}]\\ 
-Cl        & $16.863$ & $13.188$ & $6.842$ & $1521.587$ & $76.1$ & $15.8$ & $16.49$  & $13.51$  & $6.830$ &[\onlinecite{BiswasVanpoucke:JPhysChemC2013}]\\
-Br        & $16.348$ & $13.845$ & $6.862$ & $1553.001$ & $80.5$ & $23.0$ & $16.572$ & $13.619$ & $6.881$ &[\onlinecite{BiswasVanpoucke:JPhysChemC2013}]\\
-2,5F      & $17.704$ & $12.034$ & $6.832$ & $1455.568$ & $68.4$ & $12.1$ & $16.43$ & $13.034$ & $6.797$ & \\
-OH        & $16.616$ & $13.501$ & $6.852$ & $1537.046$ & $78.2$ & $2.6$  & $16.67$ & $13.492$ & $6.834$ & [\onlinecite{BiswasVanpoucke:JPhysChemC2013}]\\
-CH$_3$    & $16.687$ & $13.385$ & $6.843$ & $1528.355$ & $77.5$ & $19.4$ & $16.42$ & $13.61$  & $6.838$ & [\onlinecite{BiswasVanpoucke:JPhysChemC2013}]\\
-CF$_3$    & $16.046$ & $14.191$ & $6.883$ & $1567.271$ & $83.0$ & $29.1$ & $15.98$ & $14.212$ & $6.901$ & [\onlinecite{BiswasVanpoucke:JPhysChemC2013}]\\
-OCH$_3$   & $16.841$ & $13.197$ & $6.842$ & $1520.498$ & $76.2$ & $27.4$ & $16.222$ & $13.79$ & $6.797$ & [\onlinecite{BiswasVanpoucke:JPhysChemC2013}]\\
\end{tabular}
\begin{flushleft}
$^{a}$  Lattice parameters $a$, $b$, and $c$, as well as the volume of a conventional unit cell containing $4$ formula units ($72$--$88$ atoms) are given. The
opening angle $\theta$ and the linker rotation angle $\phi$ are defined in Fig.~\ref{fig:MIL47ref}. The starting (experimental) lattice parameters are given in
comparison.
\end{flushleft}
\end{table*}
\section{Results and Discussion}
\subsection{Crystal structure}
\begin{figure}
  \includegraphics[width=8cm,keepaspectratio=true]{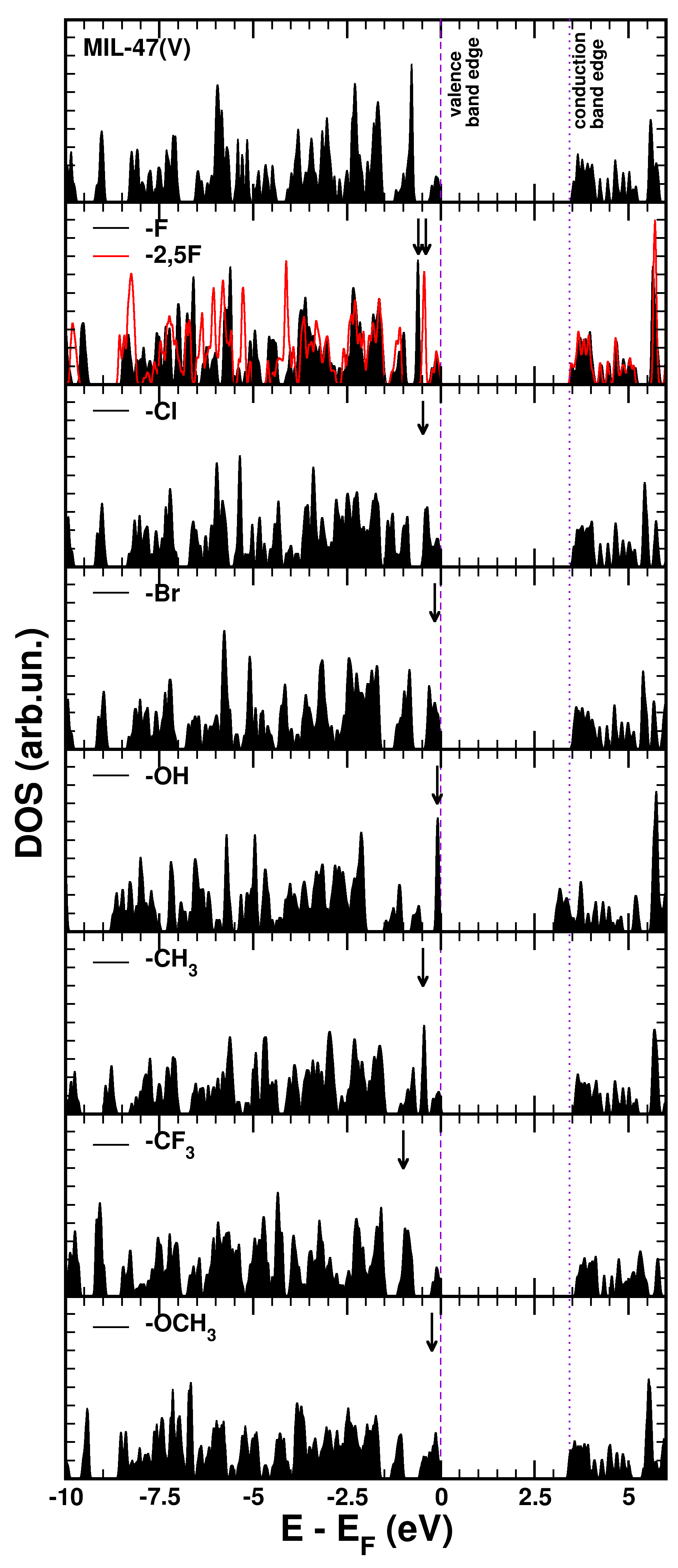}\\
  \caption{The total DOS of the MIL-47(V) and the functionalized MOFs in the AF spin configuration, obtained for the HSE06 hybrid functional. The valence and conduction band edges of the unfunctionalized MIL-47(V) AF system are indicated by the dashed and dotted lines, respectively. The position of the split-off $\pi$-band is indicated by the arrows. The V-$d$ valence band is located at the valence band edge for each system except the hydroxy-functionalized case, where it is the peak located at about $-0.6$~eV. The DOS was smoothed using a $25$~meV wide gaussian smoothing function.}\label{fig:DOSHSEAFall}
\end{figure}

\indent The structural parameters of the differently functionalized MOFs with an AF spin configuration are presented in Table~\ref{Tab:MOFStruct_AF}. Starting from the experimentally measured orthorhombic structures, the lattice angles show, upon structure optimization, only minor deviations ($\leq 1^{\circ}$) from $90^{\circ}$, and as such can be considered to remain orthorhombic. Also the short $c$-lattice vector of the functionalized MOFs, shown in Table~\ref{Tab:MOFStruct_AF}, deviates no more than $0.05$ \AA\ from its original experimental value.\cite{BiswasVanpoucke:JPhysChemC2013} The long $a$- and $b$-lattice vectors on the other hand can deviate up to $1.5$ \AA\ from the experimental values, with the $a$-lattice vector generally being longer and the $b$-lattice vector being shorter than the experimental values. As a result, the pores are narrower in comparison to the experimentally observed structure. Similar to the unfunctionalized MIL-47(V) MOF, we find that the AF and FM spin configurations have little influence on the structure (compare Table~\ref{Tab:MOFStruct_AF} and S1). For the FM configurations, the $a$-lattice vector is on average $0.126$ \AA\ longer than for the AF configuration, while the $b$- and $c$-lattice vectors are, respectively, about $0.076$ and $0.013$ \AA\ shorter. This results in an opening angle $\theta$ which is about $0.7^{\circ}$ larger for the AF configurations.\\
\indent As the system volume varies, it is meaningful to consider the opening angle $\theta$ as an indicator of the pore closure. The calculated $\theta$ for the functionalized and unfunctionalized MIL-47(V) MOF is presented in Table~\ref{Tab:MOFStruct_AF}. The results show that functionalization generally leads to a slight closing of the pore, with the \ce{-CF3} functional group being an exception.\\
\begin{figure}
  \includegraphics[width=8cm,keepaspectratio=true]{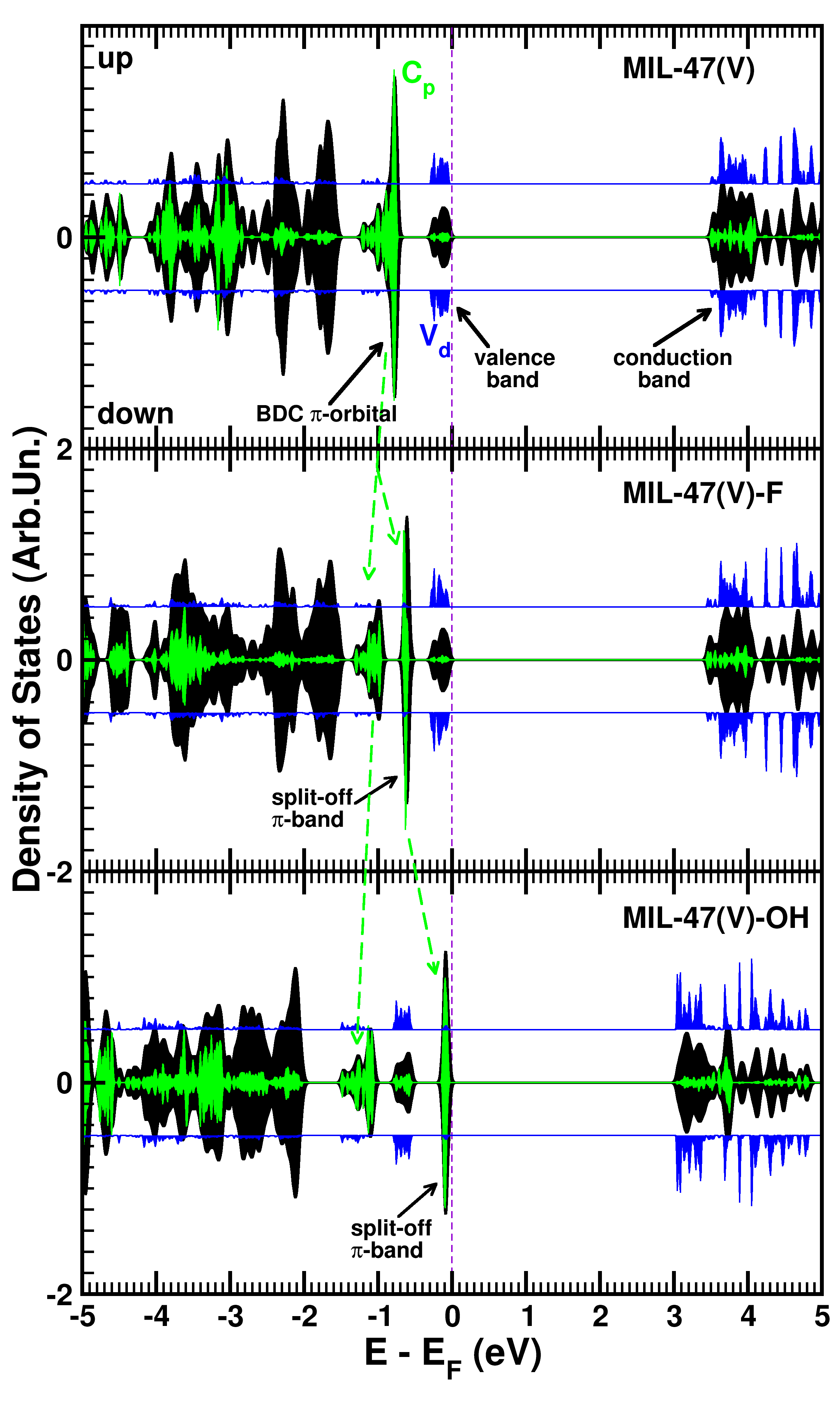}\\
  \caption{Top panel: The total DOS (black curves) for the unfunctionalized MIL-47(V) and the atom and orbital projected DOS for the V $d$-states, V$_d$, (blue off-set curves) and the C $p$-states, C$_p$, (green curves) of the C atoms in the ring of the BDC linker molecule. The V $d$-states provide the main contribution to the band gap edges, while the C $p$-states give rise to a broad band about $1$-eV below the Fermi level. This band corresponds to the $\pi$-orbital on the ring of the BDC linker. Middle panel: Same total and projected DOS states for the functionalized MIL-47(V)-F MOF. The split-off $\pi$-band is indicated. Bottom panel: Same total and projected DOS states for the functionalized MIL-47(V)-OH MOF. The split-off $\pi$-band is indicated.\\
  Green dashed arrows guide the eye with regard to the splitting of the $\pi$-band. All DOS's are obtained for the AF spin configuration using the HSE06 functional.}\label{fig:HSESOPB}
\end{figure}
\indent Since the functional groups are attached to the linkers, which can rotate freely around the bond between the central ring and the carboxylate groups, it is interesting to look at the linker rotation angle $\phi$ . As one would expect, heavier linkers lead to a larger rotation angle of the linker due to the higher moment of inertia (\textit{cf.} Table~\ref{Tab:MOFStruct_AF}): $I=\sum_{i}{m_ir_{i}^{2}}$ with $i$ the $m_i$ and $r_i$ the mass and position with regard to the rotation axis of atom $i$ of the linker, respectively. From this equation, it is clear that the internal topology of the functional group will also play an important role in determining the moment of inertia. This leads to variations of the linker rotation angle for functional groups of (almost) the same mass (\textit{e.g.} compare \ce{-F}, \ce{-OH}, and \ce{-CH3}). Comparison of the FM and AF configurations shows that the linkers consistently present a smaller rotation angle $\phi$ for the AF configurations. This difference may be an indicator of weak interaction between the functional group and the vanadyl-chains. This is corroborated by the density of states (DOS)(see~Fig.\ref{fig:HSESOPB}~and~\ref{fig:DOSHSEAFall}) which shows that a good correlation exists between the linker rotation angle $\phi$ and the overlap of the V$-d$ valence band and the split-off $\pi$-band (defined in  Fig.~\ref{fig:HSESOPB}) due to the functionalization:\cite{fn:badoverlap} (i) In case of the hydroxy-group, the smallest linker rotation is observed, and no overlap between the split-off $\pi$-orbital (arrow in Fig.~\ref{fig:DOSHSEAFall}) and the V-$d$ valence band (located at about $-0.6$~eV for this case). (ii) In case of the \ce{-Br} and \ce{-OCH3} functionalization, the V-$d$ valence band as well as the split-off $\pi$-orbital are located at the valence band edge, giving rise to full overlap. A very large linker rotation is observed for these functionalizations (see Table~\ref{Tab:MOFStruct_AF}). (iii) In case of the \ce{-Cl} and \ce{-2,5 F} functionalization, the V-$d$ valence band is again located at the valence band edge, while the split-off $\pi$-orbital (indicated with an arrow) is located just below, giving rise to partial overlap between the two. For these functionalizations an intermediate linker rotation is observed.\\
\indent The geometry of the vanadyl-chains appears to be largely unperturbed by the presence of the functional group on the linker. The chains consist of tilted, asymmetrically distorted VO$_6$ octahedra with a V=O double bond at one apex of $1.655\pm 0.005$ \AA, and a longer V$\cdots$O bond at the other apex of $2.099\pm 0.015$ \AA, similar as is found for vanadyl-acetate chains.\cite{WeeksC:JMaterChem2003} The V-V distance in the chains of the functionalized MOFs ranges from $3.431$--$3.456$ \AA, and the V$\cdots$O=V superexchange angle is $132$--$133^{\circ}$, the same as was found before for the unfunctionalized MIL-47(V), and in good agreement with the experimental value of $124^{\circ}$.\cite{VanpouckeDannyEP:2014e_Beilstein, CentroneA:Small2010} Similar as for the unfunctionalized MIL-47(V), the spin configuration has no significant influence on the vanadyl-chain geometry.\\
\subsection{Atomic charges}
\indent To quantify possible charge transfer between the functional groups and the rest of the system, we calculated the Hirshfeld-I charges for all the functionalized MIL-47(V),\cite{VanpouckeDannyEP:2013aJComputChem, VanpouckeDannyEP:2013bJComputChem, BultinckP:2007JCP_HirRef} and present them in Table~\ref{Tab:HIcharges} (PBE, AF), S2 (PBE, FM), and S3 (HSE06, AF). Since the unit cells contain four formula-units, averaged charges are presented. In most cases the standard deviation $\sigma$ on the presented charges is below $0.010$e, the exceptions being O$_{BDC}$ ($0.012<\sigma<0.016$) and C$_O$ ($\sigma<0.014$).\\
\begin{table*}[!tbh]
  \caption{Hirshfeld-I atomic charges.$^{a}$}\label{Tab:HIcharges}
  \begin{center}
  \scalebox{0.75}{
  \begin{tabular}{l|rrrrrrrrrrrrrr}
           & V      & O$_V$   & O$_{BDC}$ & C$_O$ & R      & C$_R$   & C$_1$   & C$_2$   & C$_3$   & C$_4$   & C$_5$   & H$_1$ & H$_3$ & H$_4$ \\
  \hline
MIL-47(V)  & $2.44$ & $-1.00$ & $-0.73$ & $0.77$ & $-$     & $-0.08$ & $-0.08$ & $-0.11$ & $-0.08$ & $-0.08$ & $-0.11$ & $0.12$ & $0.12$ & $0.12$ \\
-F         & $2.43$ & $-0.99$ & $-0.73$ & $0.79$ & $-0.13$ & $0.36$  & $-0.25$ & $-0.07$ & $-0.11$ & $-0.04$ & $-0.26$ & $0.15$ & $0.12$ & $0.12$ \\
-Cl        & $2.44$ & $-1.00$ & $-0.73$ & $0.78$ &  $0.04$ & $0.14$  & $-0.18$ & $-0.08$ & $-0.10$ & $-0.06$ & $-0.19$ & $0.13$ & $0.13$ & $0.11$ \\
-Br        & $2.43$ & $-1.00$ & $-0.74$ & $0.78$ &  $0.08$ & $0.10$  & $-0.16$ & $-0.09$ & $-0.10$ & $-0.07$ & $-0.17$ & $0.13$ & $0.12$ & $0.11$ \\
-2,5F      & $2.44$ & $-0.99$ & $-0.73$ & $0.80$ & $-0.13$ & $0.34$ & $-0.22$ & $-0.21$ & $-$ & $-$ & $-$ & $0.15$ & $-$ & $-$ \\
-OH        & $2.43$ & $-1.00$ & $-0.73$ & $0.79$ & $-0.10$ & $0.39$  & $-0.23$ & $-0.08$ & $-0.12$ & $-0.03$ & $-0.30$ & $0.14$ & $0.12$ & $0.12$ \\
-CH$_3$    & $2.43$ & $-1.00$ & $-0.73$ & $0.78$ & $-0.05$ & $0.20$  & $-0.15$ & $-0.10$ & $-0.10$ & $-0.07$ & $-0.18$ & $0.12$ & $0.13$ & $0.12$ \\
-CF$_3$    & $2.43$ & $-0.99$ & $-0.73$ & $0.78$ & $0.12$  & $-0.18$ & $-0.08$ & $-0.10$ & $-0.07$ & $-0.08$ & $-0.08$ & $0.12$ & $0.13$ & $0.12$ \\
-OCH$_3$   & $2.43$ & $-0.99$ & $-0.73$ & $0.78$ & $0.03$  & $0.32$  & $-0.26$ & $-0.07$ & $-0.13$ & $-0.06$ & $-0.25$ & $0.13$ & $0.12$ & $0.13$ \\
  \end{tabular}
  }
  \end{center}
\begin{flushleft}
$^{a}$ Calculated atomic charges [units: electron] for the different atoms indicated in Fig.~\ref{fig:MIL47ref}c. The PBE-electron density of the AF systems is partitioned using the iterative Hirshfeld scheme.\cite{VanpouckeDannyEP:2013aJComputChem, VanpouckeDannyEP:2013aJComputChem, BultinckP:2007JCP_HirRef} For the \ce{-2,5 F} system, the charge on a single functional group is presented, allowing for direct comparison to the \ce{-F} functionalized case.
\end{flushleft}
\end{table*}
\indent The presence of the functional group (designated as `R') only modifies the charge distribution on the ring of the BDC-linkers, while the charge on the VO$_6$ chains remains entirely unperturbed. Although all the functional groups formally have the same charge (-1), the calculated charge varies from $-0.13$ to $+0.12$e (column R in Table~\ref{Tab:HIcharges}). These rather small charges indicate that there will be very little charge transfer between the linker-ring and the functional group itself. It is interesting to note that although the functional groups differ significantly, the absolute value of the charge transfer with the ring is no larger than that of the H atoms bound to the same ring (compare R to H charges). The charge on the C$_R$ atom on the other hand varies more strongly (from $-0.18$ e up to $+0.39$ e), indicating that charge is not merely transferred towards the functional group but, more generally, away from this atom toward both the rest of the ring and the functional group. The presence of the functional group clearly breaks the symmetry of the ring, which is shown by the C and H charges. It is important to note that if multiple functional groups are present on the same linker, these groups each show charge transfer behavior as the solitary functional group (compare -F and -2,5F). For the halide series, the inductive electron withdrawing behavior of the halide atom is compensated by its lone-pair back donation as also seen in the calculated charges.\\
\indent Comparison of the obtained charges for the different spin configurations shows qualitatively and quantitatively the same picture, indicating that the influence of the functional group is localized on the linker. The qualitative picture is even retained between different levels of theory (with only mild quantitative changes).\\

\begin{table}[!tbh]
  \caption{Electronic structure related parameters.$^{a}$}\label{Tab:AFFMEnergy}
  \begin{tabular}{l|rr|rr|rr}
    & \multicolumn{2}{c}{$\Delta_{FM}^{AF}$}     & \multicolumn{2}{c}{BG$^{AF}_{eff}$}    & \multicolumn{2}{c}{BG$^{FM}_{eff}$} \\
    & \multicolumn{2}{c}{(meV/FU)} & \multicolumn{2}{c}{(eV)} & \multicolumn{2}{c}{(eV)} \\
    & PBE & HSE06 & PBE & HSE06 & PBE & HSE06\\
  \hline
MIL-47(V)  & $-68.9$ & $-11.2$ & $0.93$ & $3.43$ & $0.51 (2.50)$ & $2.98 (3.76)$ \\
\hline
-F         & $-71.9$ & $-10.3$ & $0.90$ & $3.39$ & $0.52 (2.28)$ & $2.92 (3.94)$ \\
-Cl        & $-68.5$ & $-11.7$ & $0.95$ & $3.43$ & $0.50 (2.02)$ & $2.98 (3.89)$ \\
-Br        & $-68.1$ & $-10.9$ & $0.98$ & $3.44$ & $0.51 (1.83)$ & $3.00 (3.80)$ \\
-2,5F      & $-67.6$ & $-10.5$ & $0.92$ & $3.37$ & $0.49 (2.08)$ & $2.93 (4.52)$ \\
-OH        & $-58.8$ & $-12.4$ & $0.90$ & $2.96$ & $0.46 (1.47)$ & $2.60 (3.70)$ \\
-CH$_3$    & $-65.9$ & $-11.9$ & $0.97$ & $3.47$ & $0.49 (2.22)$ & $3.04 (3.78)$ \\
-CF$_3$    & $-67.2$ & $-9.9$  & $1.00$ & $3.53$ & $0.55 (2.56)$ & $3.07 (4.72)$ \\
-OCH$_3$   & $-63.2$ & $-11.1$ & $0.98$ & $3.33$ & $0.51 (1.77)$ & $3.02 (3.96)$ \\
\end{tabular}
\begin{flushleft}
$^{a}$  $\Delta_{FM}^{AF}$ is the relative stability of the AF spin configuration compared to the FM spin configuration per formula unit (FU). BG$^{AF}_{eff}$ and BG$^{FM}_{eff}$ the effective band gap for AF and FM spin configuration, respectively.\cite{fn:effBG} For the FM systems, the band gap for the minority spin is given between brackets.
\end{flushleft}
\end{table}

\subsection{Electronic structure}
\begin{figure*}[!tb]
  \includegraphics[width=16cm,keepaspectratio=true]{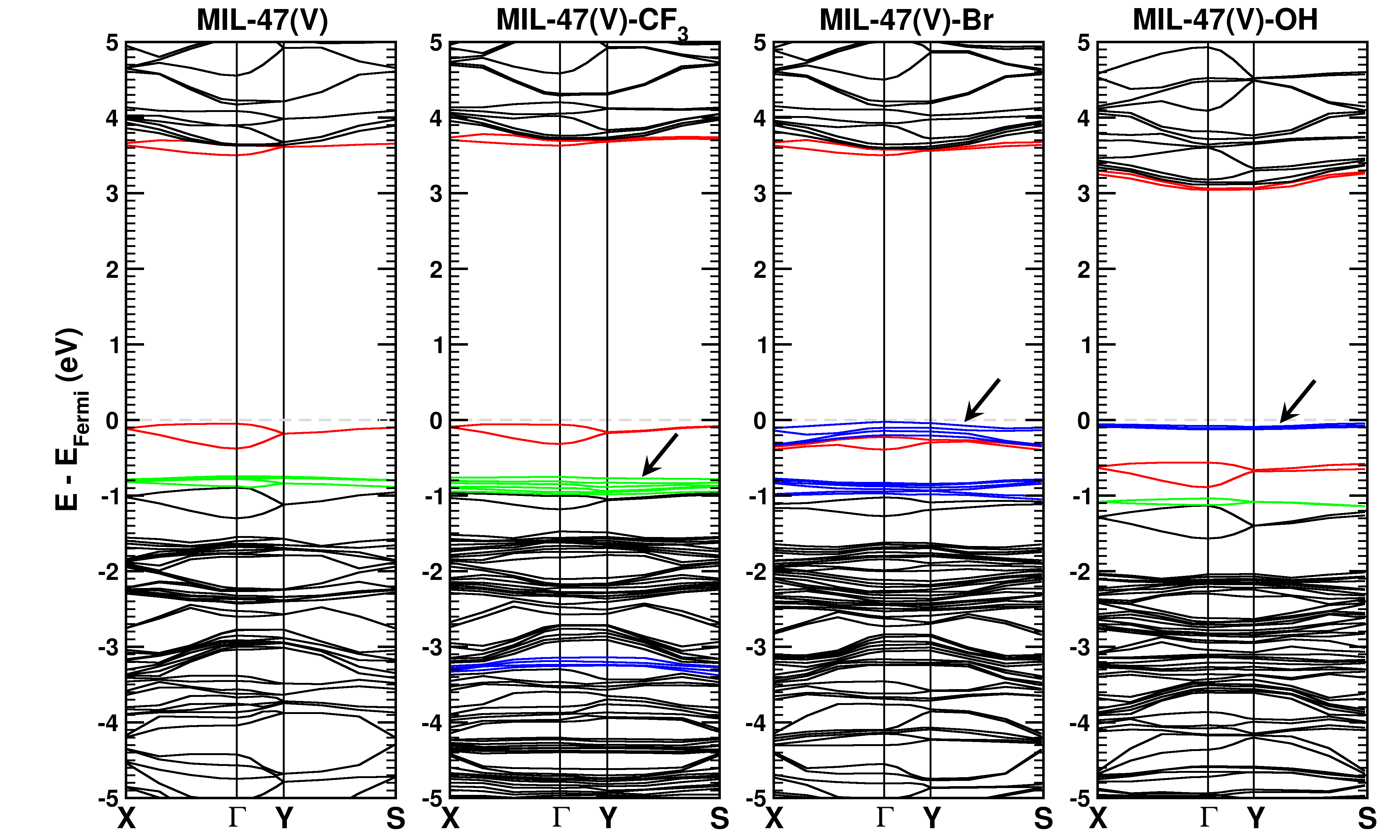}\\
  \caption{Electronic band structure along high symmetry lines for the MIL-47(V) MOF and the \ce{-CF3}, \ce{-Br}, and \ce{-OH} functionalized versions. Band structures are calculated using the hybrid HSE06 functional. Top two valence and bottom two conduction bands for the unfunctionalized MIL-47(V) are indicated in red, while top valence bands linked to the benzene ring of the BDC linkers are shown in green. Blue colored bands are those showing significant character related to the atoms of the functional group. The split-off $\pi$-bands are indicated with the arrow.
  }\label{fig:BandsHSEAF}
\end{figure*}

\indent In previous theoretical studies of the unfunctionalized MIL-47(V), it was shown that the spin configuration of the unpaired V $d$-electrons plays an important role in the physical and mechanical properties of this MOF, providing a larger contribution to the stability of the system than the geometry (large pore vs. narrow pore).\cite{VanpouckeDannyEP:2014e_Beilstein} In addition, from magnetic measurements on single crystal MIL-47(V) the ground state spin configuration is known to be AF.\cite{BartheletK:AngewChemIntEd2002} Powder-XRD studies, on the other hand, show that for powders the structure associated with a FM spin configuration provides both a qualitatively and quantitatively better fit to the experimental spectrum.\cite{BogaertsVanpoucke:CrystEngComm2015} Therefore both the FM and AF spin configurations have been considered in this work. The calculated relative stability of the AF configuration shows only little dependence on the functional group (\textit{cf.}~Table~\ref{Tab:AFFMEnergy}). However, there is a significant difference in stability when comparing the PBE and HSE06 calculations. Although for both functionals the AF configuration is clearly more stable than the FM configuration, in the case of the hybrid functional, HSE06, it is only by a mere $10$~meV/formula unit. This energy difference is comparable to the difference in ground state energy between the large and narrow pore geometries of the unfunctionalized MIL-47(V), as was shown previously.\cite{VanpouckeDannyEP:2015b_JPhysChemC} As a result, one might expect both the FM and AF spin configurations to be present in samples under ambient conditions. Of all functional groups, the hydroxy-group presents the largest modification of the stability, making the AF configuration least stable for the PBE functional, but most stable according to the HSE06 functional. This may originate from the qualitatively different electronic band structure in PBE and HSE06 case for the hydroxy-functionalized MIL-47(V), where the split-off $\pi$-orbital end up below (PBE) or above (HSE06) the V-$d$ valence band. Such a qualitative difference is not observed for the other functional groups. However, as the range of energy differences, $\Delta_{FM}^{AF}$, for the HSE06 calculations is a mere $2.5$~meV/FU, it is also important to keep in mind that this is only a factor ten larger than the selected numerical accuracy for the calculations ($1$ meV / $4$ FU).\\
\indent The calculated band gap, shown in Table \ref{Tab:AFFMEnergy}, presents the expected underestimation for the PBE functional, while the hybrid HSE06 functional gives a much larger band gap which is expected to be more in line with experiment, based on the performance of this functional for other MOFs with comparable functionalized linkers.\cite{HendrickxVanpoucke:InorgChem2015} For the unfunctionalized MIL-47(V) the obtained band gap of $3.43$~eV is comparable to the $3.63$~eV calculated by Ling and Slater.\cite{LingS:JPhysChemC2015} The small difference may originate from the difference in unit-cell volume ($1550.34$ \AA$^3$, in this work, compared to $1520.6$ \AA$^3$) and the fact that the band gap minimum is not located at the $\Gamma$-point (\textit{cf.}~Fig.~S4)\cite{VanpouckeDannyEP:2014e_Beilstein}\footnote{This requires a k-point sampling of the first Brillouin zone beyond the $\Gamma$-point only, which is know to provide poor results for the MIL-47(V).\cite{VanpouckeDannyEP:2015b_JPhysChemC}}. All functional groups, except \ce{-OH}, induce only minor changes in the effective band gap width, independent of the spin configuration. The different behaviour of the hydroxy-functionalized MIL-47(V) stems from the fact that only for this functional group, the split-off $\pi$-orbital is shifted to above the V-$d$ valence band.\\
\indent A comparison of the HSE06 DOS for the different functionalized MOFs shows qualitatively the same trend for the different spin configurations (AF in (\textit{cf.}~Fig.~\ref{fig:DOSHSEAFall}), FM in (\textit{cf.}~Fig.~S1)): The presence of a functional group on the linker leads to a splitting of the benzene $\pi$-orbital, located about $1$~eV below the fermi-level. The position split-off $\pi$-band (SOPB) is indicated with arrows in Figs.~\ref{fig:DOSHSEAFall}) and~S1. The electronic band structure for three representative functionalized systems is compared to that of the unfunctionalized MIL-47(V) in Fig.~\ref{fig:BandsHSEAF}, where the SOPB is indicated with an arrow. Similar as was observed in previous work for the functionalized UiO-66(Zr) MOF, the SOPB shifts upward for the functionalized MIL-47(V) (\textit{cf.} arrow position in Fig.~\ref{fig:DOSHSEAFall}).\cite{HendrickxVanpoucke:InorgChem2015} In contrast to the UiO-66(Zr) case, the $\pi$-band is located below the highest occupied states originating from the inorganic nodes. As a result only very large shifts of the SOPB will lead to an effective reduction of the band gap. A smaller shift, however, may result in an overlap with the valence states of the inorganic nodes, increasing the interaction between the nodes and linkers, as discussed earlier. The observed upward shift of the SOPB seems to relate well to the electron donating/withdrawing nature of the functional group, with the strongest activating functional group (\ce{-OH}) pushing the SOPB (arrow) above the V-$d$ valence band (in red) of the MIL-47(V) in Fig.~\ref{fig:BandsHSEAF}. This effectively reduces the observed band gap. Similar to what was found in the case of the functionalized UiO-66(Zr),\cite{HendrickxVanpoucke:InorgChem2015} the separation between the V-$d$ conduction and valence bands (shown in red in Fig.~\ref{fig:BandsHSEAF}) of the functionalized MIL-47(V) remains roughly unchanged. In the case of the ``weaker'' activating \ce{-OCH3} group, the SOPB fully overlaps with the V-$d$ valence band, while for the even weaker activating \ce{-CH3} and deactivating \ce{-CF3} groups the SOPB remains located below the V-$d$ valence band (\emph{cf.}~Fig.~\ref{fig:DOSHSEAFall}).\\
\indent The halide functional groups (\ce{-F}, \ce{-Cl}, and \ce{-Br}) show an effect ranging from the one found for the \ce{-CF3} to the \ce{-OCH3} functional groups, an indication that the inductive electron withdrawal from the benzene ring is being compensated to varying degrees by lone-pair back donation (\textit{cf.} halide charges in Table~\ref{Tab:HIcharges}). In the case of \ce{-F}, the inductive effect dominates leading to the SOPB remaining below the V-$d$ valence band (see Fig.~\ref{fig:DOSHSEAFall}). For the \ce{-Br} functional group, in contrast, the lone-pair back donation dominates leading to a global electron donating behavior, resulting in an SOPB located at the Fermi level (blue bands at the arrow in Fig.~\ref{fig:BandsHSEAF}). The behavior of \ce{-Cl} is intermediate between \ce{F} and \ce{Br} with the SOPB located at the lower end of the V-$d$ valence band (see Fig.~\ref{fig:DOSHSEAFall}). Furthermore, similarly as for the UiO-66(Zr),\cite{HendrickxVanpoucke:InorgChem2015} it was found that for double functionalization (\textit{cf.}~\ce{-2,5 F} in Fig.~\ref{fig:DOSHSEAFall}), the upward shift of the SOPB is increased. Following the observed trends, we expect \ce{-I} and double \ce{-Br} functionalization to give rise to an SOPB located above the Fermi-level.\\
\indent For the FM spin configuration the picture is a bit more complex, as the majority and minority spin components present a different DOS (\textit{cf.} Fig.~S1). Using the HSE06 functional, the majority spin component shows a metallic DOS (in contrast to the PBE case,\cite{fn:half-metal} which presents an insulating DOS, see Fig.~S2) while the minority spin component shows an insulating DOS. This makes the MIL-47(V) a half-metal,\cite{CoeyMD:JApplPhys2002_halfmetal} and thus an interesting material for possible spintronics applications. In addition,the role of the SOPB also becomes more explicit.\\
\indent In the case of the \ce{-OH} functionalized linkers, which presented the largest shift in the AF case, the SOPB also becomes the new valence band, located well above the, now fully occupied, majority spin V-$d$ valence band (see Fig.~S1). As a result, in contrast to the metallic DOS of the unfunctionalized FM MIL-47(V), an insulating DOS is obtained for the OH-functionalized FM MIL-47(V). For the latter the band gap is calculated to be $2.6$~eV.\\
\indent Similarly, the \ce{-OCH3} and \ce{-Br} functionalizations shift the SOPB to the top of the valence band (see Fig.~S1), transforming the metallic majority spin component into an insulator with a band gap of about $3$~eV. In the case of the other functional groups, the SOPB is located below the global Fermi level (see Fig.~S1), as for the AF spin configuration. As a result, the metallic DOS of the parent material (MIL-47(V)) is retained for the majority spin component.\\
\indent Unfortunately, the direct experimental observation of the spin filter behavior in (functionalized) FM MIL-47(V) may be rather hard, as it would require at least a single crystal sample\footnote{In a polycrystaline sample, different crystals could have opposing majority spin directions, leading to a cancelation of the effect.}. However, as the effective band gaps of the AF and FM spin configurations of all functionalized MIL-47(V) systems clearly differ ($>0.3$~eV), measurement of the band gap could provide an alternative route of determining the spin configuration experimentally. Another important point to note is that due to the very small coupling between the vanadyl chains, a system containing only FM chains but with opposing spin orientations for the neighboring chains (leading to a net magnetic moment of zero) would present the same mechanical properties as for the FM system discussed, but with an insulating electronic structure instead.\\
\indent Linking the above results to those found for functionalized BDC linkers in UiO-66(Zr) and MIL-125(Ti), one may assume similar behavior for the functionalized BDC linkers in MIL-47(V).\cite{ThorshaugKEspenFL:2013JPhysChemC, HendonCH:JACS2013, HendrickxVanpoucke:InorgChem2015} As such, it is to be expected that \ce{-NH2}, double \ce{-OH}, and single and double \ce{SH} functionalization can be used to (further) reduce the effective band gap of MIL-47(V), while \ce{-NO2} functionalization will show a behaviour similar as was seen for \ce{-OCH3} and \ce{Br}. In addition, comparison of the presented results to those of Gascon and co-workers on the MOF-5, indicates that also for double \ce{-Br} functionalization, a slight decrease of the effective band gap may be expected. In this case, the additional \ce{-Br} group is expected to push the SOPB above the MIL-47(V) Fermi-level.\cite{GasconJ:ChemSusChem2008}\\
\indent The electronic band structure along the edges of the first Brillouin zone was calculated for all systems at the PBE level (representative cases are shown in Fig.~S3(FM) and S4(AF)). Because of the extreme computational cost of hybrid functional calculations, four representative systems were selected with three high symmetry lines (\textit{cf.}~Fig.~\ref{fig:BandsHSEAF}), based on the obtained PBE results. Because the $\pi$-band is located much deeper below the V-$d$ valence band in PBE (creating a gap of $>1.5$~eV) the SOPB is much more clearly observed (\textit{cf.}~arrow positions in Fig.~S3 and S4). In the case of the halide series, the SOPB is noted to move slowly upward, with increasing atomic number, while broadening and showing an increasing dispersion (up to $0.2$~eV for Br) along directions parallel to the vanadyl chains\footnote{The presence of dispersion underscores the crystalline nature of this MOF, showing the use of standard solid-state band structure terminology to be best suited. In contrast, the often suggested terminology of HOCO and LUCO would imply fully localized molecular states with perfectly flat bands, and thus a band gap minimum located at $\Gamma$ which is not the case for the MIL-47(V)-X MOF in the AF groundstate.}. This dispersion is also visible in the HSE06 calculations for MIL-47(V)-Br (top blue curves in Fig.~\ref{fig:BandsHSEAF}) albeit somewhat hidden due to band crossings. Another interesting point to note is the fact that the dispersion of the SOPB and the V$-d$ valence have an opposite sign (see Fig.~S3 and S4), indicating that the electrons are running in opposite directions along the vanadyl chains and the pores. The same dispersion behavior is seen for the \ce{-OCH3} functionalization. In contrast, for the smaller/lighter functional groups the SOPB show much less dispersion indicating the states are more localized. A final interesting feature of the electronic band structure is the fact that the band gap of the AF ground state is not located at the $\Gamma$-point but at the U point ($\frac{1}{2}$, $0$, $\frac{1}{2}$) instead (\textit{cf.}~Fig.~S4). As a result, the (HSE06) valence band shown in Fig.~\ref{fig:BandsHSEAF} is located  slightly below the Fermi-level, at about $-55$, $-60$, $-30$, and $-80$~meV for the unfunctionalized, and the \ce{-CF3}, \ce{-Br}, and \ce{-OH} functionalized MIL-47(V), respectively. As such, often used $\Gamma$-point only calculations will overestimate the actual band gap.\\
\indent Investigation of the SOPB for \ce{-CF3} functionalized MIL-47(V) shows a behavior different from that for all other functionalisations. Unlike the other functional groups, the largest contribution to the electronic band character, due to atoms belonging to the functional group, is found in states located more than $3$~eV below the Fermi level (indicated in blue in Fig.~\ref{fig:BandsHSEAF}). In addition, the bands identified as the benzene $\pi$ band show a significant broadening, leading to the SOPB roughly retaining its position. Closer investigation of the local density of states shows a very small fraction of \ce{F} character being present in these bands, linking the broadening to the functionalization. In contrast, the \ce{-OH} functionalization leads to a very flat band with very little broadening, which is clearly linked to the O-H bond. As the SOPB is located above the V-$d$ valence band, the effective band gap is reduced (\textit{cf.}~Table~\ref{Tab:AFFMEnergy}). However, if one compares the position of the V-$d$ valence and conduction bands (indicated in red in Fig.~\ref{fig:BandsHSEAF}) in the different systems, it is clear that these bands remain in their original position, a situation reminiscent of the UiO-66(Zr) case.\\
\section{Conclusion}
\indent In this work, we investigated the influence of linker functionalization on the electronic and structural properties of the MIL-47(V) MOF. We have found that the linker functionalization induces a closing of the pore and a rotation of the linker, with heavier functional groups leading to larger linker rotations. Comparison of the systems with an anti-ferromagnetic and ferromagnetic spin configuration of the unpaired V atoms, shows a consistently smaller rotation angle for the anti-ferromagnetic systems. This may indicate that there is a weak interaction between the functional group and the vanadyl chains present, which corresponds well with the overlap of the split-off $\pi$-band and the V-$d$ valence band.\\
\indent Comparison of the atomic charges shows that the influence of the functional groups remains very localized, in line with the expected weak interaction with the vanadyl chains. The calculated electron-drawing/donation is rather small for all functional groups (comparable to H). However, the C$_R$ atom to which the functional group is bound shows a large decrease in charge, showing electrons are not only pulled toward the functional group, but also pushed onto the remainder of the benzene-ring.\\
\indent The modification of the electronic band structure of the host MIL-47(V) MOF due to the presence of functional groups is reminiscent of the behavior observed in linker functionalized UiO-66(Zr) and MIL-125(Ti) MOFs.\cite{HendrickxVanpoucke:InorgChem2015, HendonCH:JACS2013} The functional group splits the molecular $\pi$-band of the benzene ring in the BDC linkers and shifts part of it upward. This shift correlates with the electron-drawing/donating nature of the functional groups, and is modified by the presence of lone-pair back-donations (corroborated by calculated Hirshfeld-I charges) as can be seen in case of the halide series. Of all functional groups, only the hydroxy-group leads to a sufficiently large shift to result in an actual reduction of the effective band gap. In case of a ferromagnetic MIL-47(V) host, which is a half-metal, the modifications are more dramatic, as some of the functional groups give rise to a metal-insulator transition for the majority spin.\\
\indent Based on the obtained results, it is to be expected that single functionalization with \ce{-NH2} and \ce{-SH} will also give rise to a band gap reduction in the MIL-47(V) MOF (with even greater reductions upon double functionalization). From the halide groups a band gap reduction is expected for \ce{-I} and double \ce{-Br} functionalization. Furthermore, the (weak) interaction between certain functional groups and the vanadyl chains, due to a band-overlap between the V$-d$ valence band and the SOPB, might provide interesting prospects of coupling the MIL-47(V) spin configuration to the presence of guest molecules.

\section*{Acknowledgement}
\indent The author wishes to thank Shyam Biswas for providing access to his experimental data. The author is a postdoctoral researcher funded by the Foundation of Scientific Research-Flanders (FWO) (project no. 12S3415N). The computational resources and services used in this work were provided by the VSC (Flemish Supercomputer Center), funded by the Research Foundation - Flanders (FWO) and the Flemish Government -- department EWI.

\section*{Supporting Information}
The optimized atomic structures can be obtained free of charge from The Cambridge Crystallographic Data Centre: CCDC $1507811$--$1507830$. Computational details and additional structural information is available in the Supporting Information. Also calculated atomic charges and additional electronic structure results are provided. This material is available free of charge via the Internet at \url{https://pubs.acs.org/doi/10.1021/acs.jpcc.7b01491}.


\begin{thebibliography}{71}
\bibitem [{fn:({\natexlab{a}})}]{fn:badoverlap}%
  \BibitemOpen
  \href@noop {} {} ({\natexlab{a}}),\ \bibinfo {note} {the \ce{-CH3} and
  \ce{-CF3} functionalisations seem to be an exception for the observed
  correlation giving rise to larger linker rotation angles than would be
  expected from the overlap between the split-off $\pi$-orbital and V-$d$
  valence band.}\BibitemShut {Stop}%
\bibitem [{fn:({\natexlab{b}})}]{fn:effBG}%
  \BibitemOpen
  \href@noop {} {} ({\natexlab{b}}),\ \bibinfo {note} {the effective band gap
  was calculated as the difference between the maximum of the valence band and
  the minimum of the conduction band considering both the majority and minority
  electronic band structure: $max_{val.band}(minority,
  majority)-min_{cond.band}(minority, majority)$}\BibitemShut {NoStop}%
\end{thebibliography}
\end{document}